\documentclass{emulateapj}

\shorttitle{First $z>6$ quasars in DECaLS+UHS}
\shortauthors{Wang et al.}

\begin{document}

\title{First discoveries of $\lowercase{z}>6$ quasars with the DECam Legacy Survey and UKIRT Hemisphere Survey}

\author{Feige Wang\altaffilmark{1,2,3}, Xiaohui Fan\altaffilmark{2,3}, Jinyi Yang\altaffilmark{1,2,3}, Xue-Bing Wu\altaffilmark{1,3}, Qian Yang\altaffilmark{1,2,3}
            Fuyan Bian\altaffilmark{4,16}, Ian D. McGreer\altaffilmark{2}, Jiang-Tao Li\altaffilmark{5}, Zefeng Li\altaffilmark{1}, Jiani Ding\altaffilmark{2}, Arjun Dey\altaffilmark{6}, Simon Dye\altaffilmark{7},
            Joseph R. Findlay\altaffilmark{8}, Richard Green\altaffilmark{2}, David James\altaffilmark{9,10}, Linhua Jiang\altaffilmark{3}, Dustin Lang\altaffilmark{11},Andy Lawrence\altaffilmark{12},
            Adam D. Myers\altaffilmark{8}, Nicholas P. Ross\altaffilmark{12,13}, David J. Schlegel\altaffilmark{14}, Tom Shanks\altaffilmark{15}}

\altaffiltext{1}{Department of Astronomy, School of Physics, Peking University, Beijing 100871, China}
\altaffiltext{2}{Steward Observatory, University of Arizona, 933 North Cherry Avenue, Tucson, AZ 85721, USA}
\altaffiltext{3}{Kavli Institute for Astronomy and Astrophysics, Peking University, Beijing 100871, China}
\altaffiltext{4}{Research School of Astronomy and Astrophysics, Australian National University, Weston Creek, ACT 2611, Australia}
\altaffiltext{5}{Department of Astronomy, University of Michigan, 311 West Hall, 1085 S. University Ave, Ann Arbor, MI, 48109-1107, USA}
\altaffiltext{6}{National Optical Astronomy Observatory, 950 N. Cherry Ave.,Tucson, AZ 85719, USA}
\altaffiltext{7}{School of Physics and Astronomy, Nottingham University, University Park, Nottingham, NG7 2RD, UK}
\altaffiltext{8}{Department  of  Physics  and  Astronomy,  University  of  Wyoming, Laramie, WY 82071, USA}
\altaffiltext{9}{Cerro Tololo Inter-American Observatory, Casilla 603 La Serena, CHILE}
\altaffiltext{10}{Department of Astronomy, University of Washington, Box 351580, Seattle, WA 98195 , USA}
\altaffiltext{11}{Dunlap Institute for Astronomy \& Astrophysics, University of Toronto, 50 St.  George Street, Toronto, Ontario, M5S 3H4, Canada}
\altaffiltext{12}{Institute for Astronomy, University of Edinburgh, Royal Observatory, Blackford Hill, Edinburgh, EH9 3HJ, UK}
\altaffiltext{13}{Department of Physics, Drexel University, 3141 Chestnut Street, Philadelphia, PA 19104, USA}
\altaffiltext{14}{Lawrence Berkeley National Laboratory, 1 Cyclotron Road, Berkeley, CA 94720, USA}
\altaffiltext{15}{Department of Physics, Durham University, South Road, Durham DH1 3LE, UK}
\altaffiltext{16}{Stromlo Fellow}


\begin{abstract}
We present the first discoveries from a survey of $z\gtrsim6$ quasars using imaging data from the DECam Legacy Survey (DECaLS) in the optical, the UKIRT Deep Infrared Sky Survey (UKIDSS) and a preliminary version of the UKIRT Hemisphere Survey (UHS) in the near-IR, and ALLWISE in the mid-IR. 
DECaLS will image 9000\,deg$^2$ of sky down to $z_{\rm AB}\sim23.0$, and UKIDSS and UHS, which will map the northern sky at $0<DEC<+60^{\circ}$, reaching $J_{\rm VEGA}\sim19.6$ (5-$\sigma$). The combination of these datasets allows us to discover quasars at redshift $z\gtrsim7$ and to conduct a complete census of the faint quasar population at $z\gtrsim6$. In this paper, we report on the selection method of our search, and on the initial discoveries of two new, faint $z\gtrsim6$ quasars and one new $z=6.63$ quasar in our pilot spectroscopic observations. The two new $z\sim6$ quasars are at $z=6.07$ and $z=6.17$ with absolute magnitudes at rest-frame wavelength 1450\,\AA\ being $M_{1450}=-25.83$ and $M_{1450}=-25.76$, respectively. These discoveries suggest that we can find quasars close to or fainter than the break magnitude of the Quasar Luminosity Function (QLF) at $z\gtrsim6$. The new $z=6.63$ quasar has an absolute magnitude of $M_{1450}=-25.95$. This demonstrates the potential of using the combined DECaLS and UKIDSS/UHS datasets to find $z\gtrsim7$ quasars. 
Extrapolating from previous QLF measurements, we predict that these combined datasets will yield  $\sim200$ $z\sim6$ quasars to $z_{\rm AB} < 21.5$, $\sim1{,}000$ $z\sim6$ quasars to $z_{\rm AB}<23$, and $\sim 30$ quasars at $z>6.5$ to $J_{\rm VEGA}<19.5$. 
\end{abstract}


\section{Introduction}

The first billion years after the Big Bang is a crucial epoch for understanding the history of cosmic reionization, and how the origins of the super-massive black hole (SMBH) population are linked to galaxy formation and evolution. High redshift quasars are the most promising tracers with which to address these issues \citep[e.g.][]{fan06a}. However, they have a very low spatial density and their selection is heavily contaminated by cool dwarfs.

Over the last fifteen years, more than 100 quasars at $z\gtrsim5.7$ have been discovered from several optical surveys such as the Sloan Digital Sky Survey \citep[SDSS;][]{york00} searches \citep[][]{fan00,fan01a,fan03,fan04,fan06b,jiang08,jiang09,jiang15,jiang16}, the Canada$-$France$-$Hawaii Telescope High-$z$ Quasar Survey \citep[CFHQS;][]{willott07,willott09,willott10}, the Pan-STARRS1 distant quasar survey \citep{morganson12,banados14,banados16}, the ATLAS \citep{carnall15} and Dark Energy Survey quasar searches \citep{reed15}, and the Subaru High-$z$ Exploration of Low-Luminosity Quasars survey \citep[SHELLQs;][]{matsuoka16}. 

Recently, several groups have significantly improved the efficiency for selecting $z\sim6$ quasars \citep[e.g.][]{wu15, carnall15, wang16} by combining optical and {\it Wide-field Infrared Survey Explorer} \citep[WISE;][]{wright10} colors. In addition, the advent of several large area near-IR sky surveys has prompted the discovery of a handful of $z>6.5$ quasars \citep{venemans13,venemans15}, with the highest redshift at $z$ = 7.1 \citep{mortlock11}.  Several key results have emerged from studies of these high-$z$ quasars. One example is the discovery of complete Gunn-Peterson absorption troughs and Gunn-Peterson damping wings from an increasingly neutral intergalactic medium (IGM), which mark the end of the reionization epoch at $z > 6$ \citep[e.g.][]{becker01,fan06a,mortlock11}. Another example is the discovery of mature quasars powered by billion $\rm M_\odot$ black holes at $z > 6$ \citep[e.g.][]{mortlock11,wu15}, which present significant challenges to theories of early supermassive black hole formation \citep[e.g.][]{pacucci15}. Even though, constructing a larger complete census of quasar sample at $z\gtrsim6$ is still needed to put stronger constraints on the formation of earliest black holes and the history of reionization.

In this paper, we report the discoveries of $z\gtrsim6$ quasars from a new survey utilizing a number of optical and near-IR datasets. The DECam Legacy Survey (DECaLS)\footnote{\url{http://legacysurvey.org/}} is a three-pass public survey that will image 9,000\,deg$^2$ of sky in the region $-20 \lesssim Decl. \lesssim +32 ^{\circ}$ to ($AB$ 5-$\sigma$ point-source) depths of $g\sim 24.7$, $r \sim 23.9$, and $z\sim23.0$. DECaLS, together with the $g,~r-$bands Beijing-Arizona Sky Survey \citep[BASS\footnote{\url{http://batc.bao.ac.cn/BASS/doku.php}};][]{zou17} and the Mayall $z$-band Legacy Survey (MzLS)\footnote{\url{http://ast.noao.edu/data/surveys}} provide imaging to support target selection for the Dark Energy Spectroscopic Instrument (DESI)\footnote{\url{http://desi.lbl.gov/}}, and are collectively referred to as the DESI Legacy imaging survey (DELS; Dey et al. 2017 in preparation). 
With large survey coverage and depths that reach 1.5--2.5\,mag beyond the SDSS, these new surveys provide significant potential for selecting high redshift quasars. 
In addition, we use data from a preliminary version of the UKIRT Hemisphere Survey (UHS). This is a J-band survey of the northern sky ($Decl. <60^\circ$) to a depth of $J=19.6$, supplementing the area already covered by UKIDSS. The survey was begun by the UKIDSS consortium, but is being completed by the new operators of UKIRT, University of Arizona and University of Hawaii.  The data is initially proprietary but is intended to be public in due course, through the same interface as UKIDSS, i.e. the WFCAM Science Archive\footnote{\url{http://surveys.roe.ac.uk/wsa/}}. The survey is briefly described by \cite{lawrence13} and will be fully reported in Dye et al. (2017 in preparation).
The combination of DELS and UHS potentially allows quasars to be detected to $z\gtrsim7$ over a large area ($\sim 12{,}000$\,deg$^{2}$) at high Galactic latitude.

The main goals of this paper are to present our basic method for selecting $z\gtrsim6$ quasars and the discoveries from our pilot spectroscopic survey. More complete statistical results will be presented in future work using larger samples of confirmed high-$z$ quasars. Throughout the paper, optical magnitudes are reported on the AB system, near-IR and mid-IR magnitudes on the Vega system. We adopt a standard $\Lambda$CDM cosmology with Hubble constant $H_0=70\,{\rm km~s}^{-1}\,{\rm Mpc}^{-1}$, and density parameters $\Omega_{\rm M}=0.3$ and $\Omega_{\Lambda}=0.7$.

\section{IMAGING DATA}

Our quasar selection is based on photometry from deep optical, and near- and mid-IR imaging surveys. 

For $grz$-band fluxes, we use DECaLS Data Release 2 (DR2). It includes DECam data from $z$-band observations in August 2013 (NOAO program 2013A-0741) and $grz$-band observations from August 2014 through June 2015 (NOAO program 2014B-0404). It also includes publicly released data from the Dark Energy Survey that lie within the DESI footprint.
In total, the DR2 data cover 2078\,deg$^{2}$ in $g$, 2141\,deg$^{2}$ in $r$ and 5322\,deg$^{2}$ in $z$, of which 1807\,deg$^{2}$ has been observed in all three filters. Source extraction is conducted by applying the {\em Tractor}\footnote{\url{ https://github.com/dstndstn/tractor}} to the calibrated images. The {\em Tractor} is a code for inference modeling of astronomical objects and performing source extraction on pixel-level imaging data (Lang et al. 2017 in prep). The {\em Tractor} takes the individual images as input from multiple exposures in multiple bands with different seeing. A simultaneous fit is performed for sources to the pixel-level data of all images. This produces object fluxes and colors that are consistently-measured across the wide-area imaging surveys.
The 5-$\sigma$ point source depths for the area in DR2 with only one pass are $g=23.61$, $r=23.20$ and $z=22.17$.

At near-IR wavelengths, we combine public data from the UKIDSS Large Area Survey \citep[ULAS;][]{lawrence07}, VISTA Kilo-degree Infrared Galaxy Survey \citep[VIKING;][]{arnaboldi07} and VISTA Hemisphere Survey \citep[VHS;][]{mcmahon13}, with data from a proprietary release within the UHS consortium (Dye et al. 2017 in preparation). These near-IR datasets cover a total area of approximately 14{,}000\,deg$^2$ with J-band observations and include 702\,million sources at $Decl. >-10^{\circ}$.

\begin{deluxetable}{ccccc}
\tabletypesize{\scriptsize}
\tablecaption{Photometric Information of Datasets we used in this paper\label{tbl-1}}
\tablewidth{0pt}
\tablehead{
\colhead{Survey} & \colhead{Band} & \colhead{Depth (5-$\sigma$) } & \colhead{AB offset} & \colhead{Ref} 
}
\startdata
SDSS & $i$ & 22.5 & 0.0\tablenotemark{a} & \cite{gunn98} \\
DECaLS & $g$ &23.6\tablenotemark{b} & 0.0 & Lang et al. in prep\\
DECaLS & $r$ &23.2\tablenotemark{b} & 0.0 &Lang et al. in prep\\
DECaLS & $z$ & 22.2\tablenotemark{b}& 0.0 & Lang et al. in prep\\
UHS & $J$ & 20.5& 0.938 & Dye et al. in prep\\
ULAS & $J$ & 20.5& 0.938 & \cite{lawrence07}\\
VHS & $J$ & 20.9& 0.937 &\cite{mcmahon13}\\
VIKING & $J$ & 22.1& 0.937 &\cite{arnaboldi07}\\
ALLWISE & $W1$ &20.3\tablenotemark{c} & 2.699& \cite{wright10}
\enddata
\tablecomments{All magnitudes in this table are in AB system.}
 \tablenotetext{a}{http://www.sdss.org/dr12/algorithms/fluxcal/}
 \tablenotetext{b}{The magnitude depth here is for those area with only one photometric pass.}
 \tablenotetext{c}{This value was estimated using the magnitude-error relation from \cite{yang16} using the number of coverage equals to 24, which corresponds to the mean coverage of ALLWISE dataset.}
 
\end{deluxetable}

\begin{deluxetable*}{ccrrrrrrrrrrrrrrcrl}
\tabletypesize{\scriptsize}
\tablecaption{Previously Known Quasars in the DECaLS DR2 Regions that satisfy our selection method \label{tbl-1}}
\tablewidth{0pt}
\tablehead{
\colhead{Name} & \colhead{Redshift} &
\colhead{$i_{\rm AB}$} & \colhead{$z_{\rm AB}$} & \colhead{$z_{\rm AB} ~({\rm DECaLS})$} & \colhead{$J$} & \colhead{$W1$} &\colhead{Ref} 
}
\startdata
J000552.34$-$000655.8  & 5.85 & $22.92\pm0.36$ & $20.45\pm0.17$ & $20.85\pm0.06$ & $19.75\pm0.19$ & $17.30\pm0.16$ & \cite{fan04}\\
J083643.85$+$005453.3 & 5.82 & $21.06\pm0.09$ & $18.81\pm0.05$ & $18.96\pm0.01$ & $17.70\pm0.01$ & $15.30\pm0.04$ & \cite{fan01a}\\
J083955.36$+$001554.2 & 5.84 & $23.46\pm0.33$ & $21.09\pm0.05$ & $21.36\pm0.10$ & $19.88\pm0.10$ & $17.46\pm0.18$ &\cite{venemans15}\\
J141111.29$+$121737.4 & 5.93 & $23.56\pm0.37$ & $19.61\pm0.07$ & $19.81\pm0.02$ & $19.20\pm0.09$ & $16.59\pm0.07$  &\cite{fan04}\\
J231546.57$-$002358.1 & 6.117& $24.90\pm0.28$ & $20.88\pm0.08$ & $20.77\pm0.07$ & $19.94\pm0.08$ & $17.56\pm0.21$  &\cite{jiang08}
\enddata
\end{deluxetable*}

At mid-IR wavelengths, we use the ALLWISE release\footnote{\url{http://wise2.ipac.caltech.edu/docs/release/allwise/}} of the WISE data, which combines  the original WISE survey \citep{wright10} with the NEOWISE \citep{mainzer11} post-cryogenic phase.
ALLWISE covers the entire mid-infrared sky and includes 747\,million sources\footnote{\url{http://wise2.ipac.caltech.edu/docs/release/allwise/}} with enhanced photometric sensitivity and accuracy and improved astrometric precision compared to the (previous) WISE All-Sky Data Release.

\section{Target Selection and Spectroscopic Observations}

\subsection{$z\sim6$ Quasar Candidate Selection}

At $z\sim6$, most quasars are not detected in the $g$- and $r$-bands because of the presence of Lyman limit systems (LLSs), which are optically thick to the continuum radiation from quasars. Meanwhile the Lyman series line absorptions begin to dominate in the $i$-band and $\rm Ly\alpha$ emission moves to the $z$-band \citep{fan01a}, thus the $i$-dropout technique is the most effective method for selecting $z\sim6$ quasars. However, there is no deep $i$-band photometry from DECaLS, necessitating the use of a $g$- and $r$-band dropout technique to select preliminary candidates. We initially selected high-$z$ quasar candidates from point sources (type="PSF") in DECaLS DR2 using Equations.\ 1--3. We cross-matched these objects with the IR-survey data described in \S2 using a radius of 2\farcs0. We further rejected objects having a near-IR probability of being a galaxy (pGalaxy) of pGalaxy$> 0.95$ \citep{venemans13}. 

Redwards of $\rm Ly\alpha$, quasar SEDs can typically be described by a power-law, which descends less steeply from $J$-band to $W1$ than is typical for late M-dwarf stars. Consequently, quasars are redder in $J-W1$ than M dwarfs. The $\rm Ly\alpha$ emission of $z\lesssim6.5$ quasars is located in the $z$-band, meaning that quasars have bluer $z-J$ colors than is typical for L dwarfs. For these reasons, we further use Equations.\ 4--5 to select $z\lesssim6.5$ quasar candidates. 

There are 17 previously known $z\sim6$ quasars that have DECaLS $z$-band detections, near-IR $J$-band detections and ALLWISE detections, and most of them are included in our color selection box (blue-dashed-line bounds of Figure~1). The five such quasars that have observations in all of $g$, $r$ and $z$ in DECaLS DR2 are listed in Table~1.

As DECaLS does not observe in the $i$-band, we cross-matched our candidates with SDSS DR12 sources using a radius of 2\farcs0 and removed targets with $i-z<2.0$ or $S/N(i)>3.0$ (Equation.\ 6). In order to further reject false $i$-dropouts, we performed forced aperture photometry for candidates in SDSS calibrated $i$-band images and rejected those targets that did not have SDSS photometric observations, or that had $i_{\rm APER}-z<2.0$. Since photometric uncertainties depend on aperture size, we adopted $4''$ ($\sim$3 times the PSF FWHM) aperture photometry in order to select high redshift quasars. This choice represents a tradeoff between reducing the aperture correction and reducing background noise. Remarkably, all five $z\sim6$ quasars listed in Table~1 survive all of our selection cuts. After removing those five known quasars, we selected 251 $z\sim6$ quasar candidates in total.

\begin{deluxetable*}{ccrrrrrrrrrrrrrrcrl}
\tabletypesize{\scriptsize}
\tablecaption{Properties of three new quasars reported in this paper. \label{tbl-2}}
\tablewidth{0pt}
\tablehead{
\colhead{Name} & \colhead{Redshift} & \colhead{$m_{1450}$} & \colhead{$M_{1450}$} &
\colhead{$i_{\rm AB}$} & \colhead{$z_{\rm AB} ~(DECaLS)$} & \colhead{$J$} & \colhead{$W1$} & \colhead{J-band Survey}
}
\startdata
DELS J121721.35$+$013142.52  & 6.17$\pm$0.05 &  20.93 & -25.76  &  $>23.5$ & $20.89\pm0.04$ & $20.21\pm0.16$ & $17.58\pm0.21$ & VIKING\\
DELS J155909.09$+$221214.43 & 6.07$\pm$0.02 &  20.88 & -25.83  &   $>23.5$ & $20.60\pm0.05$ & $19.86\pm0.25$ & $17.14\pm0.11$ & UHS\\
DELS J104819.09$-$010940.21 & 6.63$\pm0.08$ & 20.90& -25.95&    $>23.5$ & $21.95\pm0.08$ & $19.71\pm0.17$ & $17.34\pm0.17$ & ULAS
\enddata
\end{deluxetable*}

\begin{figure}[tbh]
\centering
\includegraphics[width=0.5\textwidth]{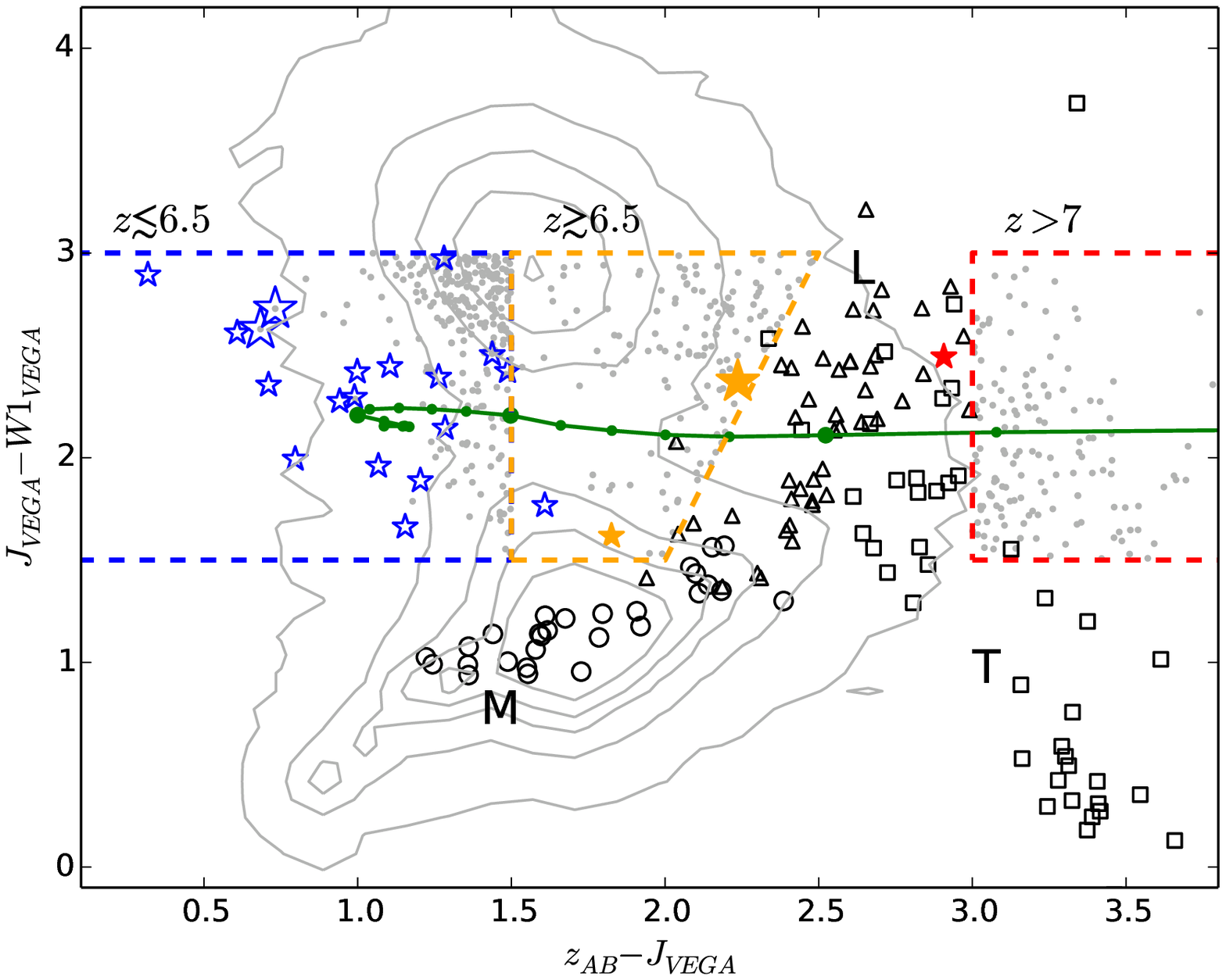}
\caption{The $z-J$ vs. $J-W1$ color-color diagram. The green line and green filled circles represent the color-$z$ relation predicted using simulated quasars \citep{mcgreer13,yang16} from $z=5.5$ to $z=7.1$, in steps of $\Delta z =0.1$. The large green circles highlight the colors at $z=6.0$, 6.5 and 7.0. The smaller blue, orange and red asterisks depict previously known $5.5 \le z <6.5$ quasars, $z\ge 6.5$ quasars and $z=7.1$ quasars, respectively. The larger blue and orange asterisks represent our newly discovered  $z\sim6$ and $z\sim6.6$ quasars. The black circles, triangles, and squares depict the positions of M, L, and T dwarfs, respectively \citep{kirkpatrick11,best15}. The horizontal and vertical dashed lines depict the broad color cuts we use to select quasars at $z\lesssim6.5$ (blue), $z\gtrsim6.5$ (orange) and $z > 7$ (red). The grey contours show the density distribution of actual data and the small grey points indicate selected candidates at each color box.}
\end{figure}

Our criteria for targeting $z\sim6$ quasars are summarized as:
\begin{equation}
S/N(g)<3.0
\end{equation}
\begin{equation}
r-z > 3.0~{\rm or}~S/N(r)<3.0
\end{equation}
\begin{equation}
z<22.0
\end{equation}
\begin{equation}
z-J<1.5
\end{equation}
\begin{equation}
1.5<J-W1<3.0
\end{equation}
\begin{equation}
i-z > 2.0~{\rm or}~S/N(i)<3.0
\end{equation}

\subsection{$6.5\lesssim z \lesssim7$ Quasar Candidate Selection}
We present our selection criteria for $6.5\lesssim z \lesssim7$ quasar candidates in Equations.\ 7--11. In addition to these equations, as was the case for our $z\lesssim6.5$ candidates, we further enforced pGalaxy$< 0.95$ and type="PSF" in order to target point sources. Primarily, we used the $z-J/J-W1$ color-color cuts (Equations.\ 7--8 and the region bounded by the orange dashed line in Figure 1) to select  $6.5\lesssim z \lesssim7$ quasar candidates. To further remove L dwarfs, which are serious contaminants of quasar samples at these redshifts, we included a cut on $Y-J$ color. The pilot sample of $z\sim6.5$ quasar candidates that we discuss in this work was selected from the ULAS DR10 spring sky imaging ($100^\circ < \alpha < 300^\circ$). The final quasar selection will be mainly focused on the overlapped region of DECaLS and UHS area, where the $Y$ band is not available, so we will not include the Equations.\ 10--11 in our future work.
We cross-matched the resulting candidates with SDSS DR12 imaging sources, and with ALLWISE, using a 2\farcs0 radius. Typically, we do not expect $z\gtrsim6.5$ quasars to be detected in the relatively shallow SDSS $z$-band. Thus, we retained objects that are not detected in the SDSS but do have ALLWISE counterparts and cross-matched them with DECaLS DR2. 
Our criteria for targeting $6.5 \lesssim z \lesssim 7$ quasars, which results in 132 candidates, are summarized as:

\begin{equation}
1.5<J-W1<3.0
\end{equation}
\begin{equation}
J-W1>3(z-J-1.5)
\end{equation}
\begin{equation}
\sigma(J)<0.2~{\rm and}~z-J>1.5
\end{equation}
\begin{equation}
\sigma(Y)<0.2
\end{equation}
\begin{equation}
z-Y>1.5~{\rm and}~Y-J<0.8
\end{equation}


\subsection{$z \gtrsim7$ Quasar Candidate Selection}
Finally, we conducted searches for $z \gtrsim 7$ quasar candidates in areas where DECaLS and UHS currently overlap. 
We started from the UHS spring sky data (100$^\circ< \alpha <300^\circ$) and rejected sources that were detected in SDSS DR12  or that were undetected in ALLWISE. We then cross-matched these initial candidates with DECaLS DR2 and used Equations.\ 12--14 together with pGalaxy $< 0.95$ to select $z \gtrsim 7$ candidates. Such 149 candidates are bounded by the red dashed lines in Figure 1, and their selections are summarized by:

\begin{equation}
\sigma(J)<0.2
\end{equation}
\begin{equation}
1.5<J-W1<3.0
\end{equation}
\begin{equation}
z-J>3.0
\end{equation}

\subsection{Selection Function}
We use simulations to estimate the preliminary selection function of our criteria, i.e., the color cuts and survey limits that we applied in \S3.1-\S3.3. We generate a grid of model quasars using the simulations by \cite{yang16}, which is an updated version of the simulations by \cite{mcgreer13}. The model quasar spectra are designed to reproduce the colors of SDSS BOSS quasars in the range $2.2<z<3.5$ \citep{ross12}. Each quasar spectrum consists a broken power-law continuum, a series of emission lines with Gaussian profiles, and Fe emission templates. The spectra also involved absorptions from neutral H absorption in $\rm Ly{\alpha}$ forests. 
The final photometry of simulated quasars are derived from the spectra models and photometric errors appropriate for each survey. We obtain the relation between magnitude and error in the $z$ and $J$ bands by obtaining a large representative sample of objects from those surveys. As the DECaLS, and near-IR surveys we used haven't reach the final uniform depth as listed in Table 1, the magnitude-error relation we used here can only represent an average effect of the current status of each survey. The more detailed analysis of the magnitude-error relation will be investigated in the future work, i.e., the quasar luminosity function estimations with a larger quasar sample and final uniform datasets. The purpose for investigating the preliminary selection function here is to present an rough idea about which fraction of quasars we are missing due to our color criteria. Figure 2 gives an example of the selection function for our $z\sim6$, $z\gtrsim6.5$ and $z\gtrsim7$ color/limit criteria in the overlapped region of DECaLS and UHS+ULAS. There is almost no difference in the selection function for VHS and VIKING surveys except the quasar missing fractions at fainter end is lower for VIKING surveys because the VIKING survey is much deeper than others. As we discussed in \S3.2, the Y band will not be available in the overlapped region of DECaLS and UHS, we do not include the $Y-J$ and $z-Y$ colors when estimate the selection function of $z\gtrsim6.5$ criteria. Figure 2 demonstrates that DECaLS together with near-IR surveys like UHS will enable us to search high redshift quasars with a wide redshift range, i.e., $5.7\lesssim z \lesssim 6.9$ and $z\gtrsim7.2$. 

\begin{figure*}[tbh]
\centering
\includegraphics[width=1.0\textwidth]{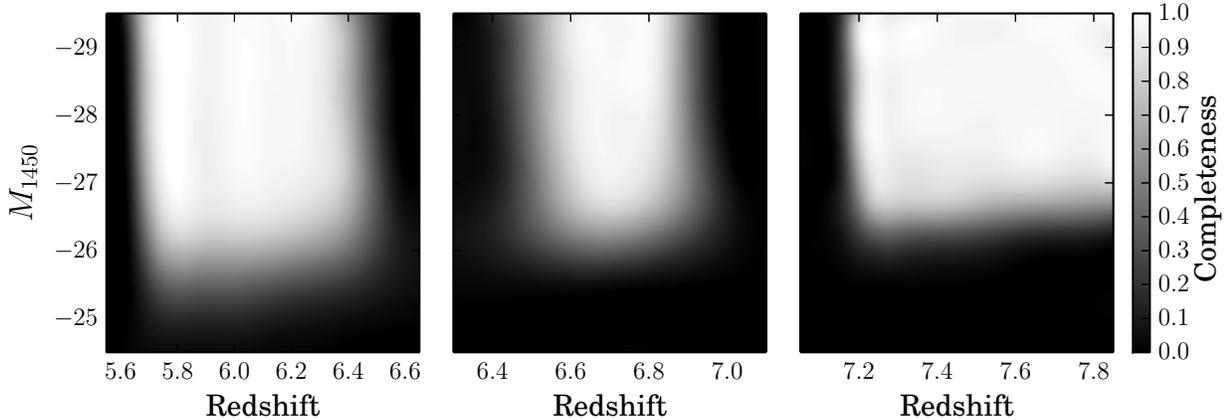}
\caption{Selection function of our selection criteria in the overlapped region of DECaLS and UHS+ULAS. The selection function for $z\sim6$, $z\gtrsim6.5$ and $z\gtrsim7$ criteria are shown in the left panel, middle panel and right panel, respectively.}
\end{figure*}

\subsection{Spectroscopic Observations}
We conducted optical spectroscopy of our $z\sim 6$ quasar candidates using the 6.5m MMT telescope in the U.S.\ and the 2.3m ANU telescope in Australia. We observed two $z\sim6$ quasar candidates with relatively high priorities in Spring 2016. High priority, in this context, means blue in $z-J$ and far from the L dwarf locus in the $z-J/J-W1$ plane. Additional candidates that satisfy our selection criteria will be further investigated as part of future work using larger DECaLS data releases.

Spectra of DELS J121721.35$+$013142.52\footnote{The naming convention of objects in the DESI Legacy Survey area (DECaLS, MzLS, BASS) is DELS Jrrmmss.ss+ddmmss.s, where RA and DEC are given in J2000.0 coordinates.} (hereafter J1217$+$0131) were obtained on the night of 2016 April 13 (UT) using the Wide Field Spectrograph \citep[WiFeS;][]{dopita07}, an integral-field double-beam image-slicing spectrograph on the ANU 2.3m Telescope at Siding Spring Observatory. We obtained three 20\,min exposures with the R3000 grating, which has a resolution of $R=3000$ and covers wavelengths between 5300\,\AA\ and 9800\,\AA.

Spectra of DELS J155909.09$+$221214.43 (hereafter J1559$+$2212) were obtained using the Red Channel spectrograph \citep{schmidt89} on the MMT telescope. Two 30\,min exposures were taken on 2016 May 29 (UT). We used the 
270\,$\rm l\,mm^{-1}$ grating centered at 8500\,\AA, providing coverage from 6600\,\AA\ to 1.0\,$\mu$m. We used a $1\farcs0$ slit, which provided a resolution of $R\sim 640$. 

The main spectral features of $z\gtrsim6.5$ quasars are redshifted beyond 9000\,\AA. We therefore observed our $z$-dropouts with the Folded-Port Infrared Echellette \citep[FIRE;][]{simcoe08}, an IR echelle/longslit spectrograph on the 6.5\,m Magellan/Baade Telescope. Observations were conducted in April 2016 using longslit mode with 5--10\,min exposures. The observational setup provided a resolution of $R\sim300$--500 from $K$-band to $J$-band. In total, we observed 23 $z\gtrsim6.5$ candidates and 19 $z>7$ candidates.

The spectra taken using MMT/Red were reduced using standard IRAF routines. The WiFeS data were reduced using a Python-based pipeline called PyWiFeS \citep{childress14}. 
The spectra taken by Magellan/FIRE were reduced using a custom set of Python routines written by one of us (F.W.). 
Flux calibrations for all spectra were calculated using standard star observations obtained on the same night as the science exposures and were scaled to $z$- or $J$-band magnitudes in order to produce absolute flux calibrations. The spectra were de-reddened to account for Galactic extinction using a Milky Way reddening law \citep{cardelli89} and the Galactic dust map of \citet{sf11}.

\begin{figure}[tbh]
\centering
\includegraphics[width=0.5\textwidth]{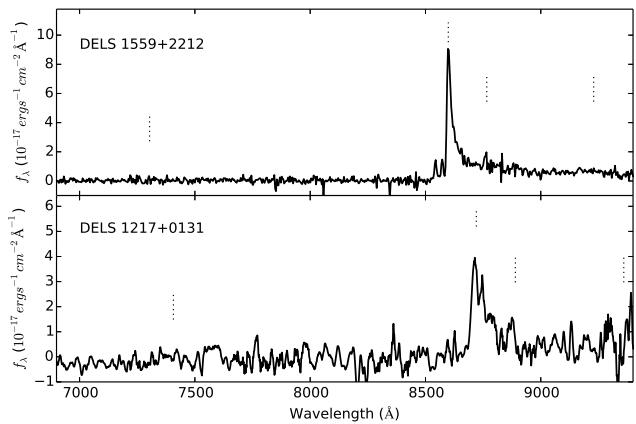}
\caption{Spectra of two newly discovered $z\sim6$ quasars. The upper panel shows J1559$+$2212 at redshift $z=6.07$ and the lower panel shows J1217$+$0131 at redshift $z=6.17$. The dotted lines mark the expected positions of four emission lines: $\rm Ly\beta$, $\rm Ly\alpha$, N\,{\sc v} and O\,{\sc i} from left to right.}
\end{figure}

\section{Results}

\begin{figure}[tbh]
\centering
\includegraphics[width=0.5\textwidth]{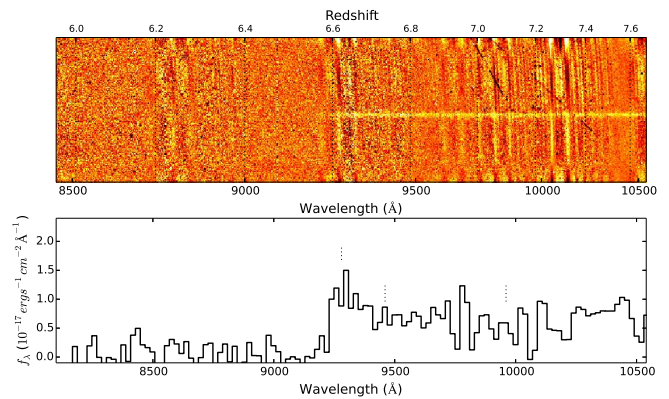}
\caption{Low resolution near-IR 2D and 1D spectra (background-subtracted and flux-calibrated) of J1048-0109. There is essentially zero flux bluewards of 
9140\,\AA, implying that J1048-0109 is a quasar at $z\sim6.6$. Note that the 2D spectrum is plotted at the native pixel-scale but the 1D spectrum has been re-binned to 20\,\AA\,pixel$^{-1}$.}
\end{figure}

We identify both of the two $z\sim6$ quasar candidates with optical spectroscopy to be $z>6$ quasars; J1217$+$0131, at $z=6.17$ and J1559$+$2212, at $z=6.07$. 
The redshifts were determined by using the visual recognition software ASERA \citep{yuan13}. ASERA is an interactive semi-automated toolkit that allows the user to visualize spectra and to estimate a redshift by fitting the  
spectrum to the SDSS quasar template of \citet{vanden01}. For the spectra of J1217$+$0131 and J1559$+$2212, we
used ASERA to visually fit the $\rm Ly \alpha$ and N\,{\sc v} emission lines.
Figure 2 shows the 1-D spectra of these two newly discovered $z\sim6$ quasars. In Table 2, columns $m_{1450}$ and $M_{1450}$ list the apparent and absolute AB magnitudes, respectively, of the continuum of these quasars
at rest-frame 1450\,\AA. These magnitudes were calculated by fitting a power-law continuum $f_\nu \sim \nu^{-0.5}$ \citep{vanden01} to the spectrum of each quasar. The power-law continuum was then normalized to match visually identified continuum windows that contain minimal contribution from quasar emission lines and from sky OH lines. We measure 
the absolute magnitudes of J1217$+$0131 and J1559$+$2212 to be -25.76 and -25.83, respectively. 

J1217$+$0131 is the first $z\sim6$ quasar discovered using DECaLS imaging. It was also independently discovered by \cite{banados16} using Pan-STARRS1 imaging. 

J1559$+$2212 has a very strong (but narrow) $\rm Ly \alpha$ emission line with a rest-frame equivalent width of 102\,\AA\ and a full width at half maximum (FWHM) of 4\,\AA. These values can be  compared to the mean values for typical $z\sim6$ quasars of 35\,\AA\ \citep{banados16} and 10\,\AA\ \citep[e.g.][]{jiang08}, respectively. J1559$+$2212  is the first $z\sim6$ quasar discovered using a combination of DECaLS and UHS data. We expect to discover more $z\sim6$ quasars in future work as the overlap between DECaLS and UHS increases producing new regions that are covered, for the first time, by both deep optical and near-IR imaging.

\begin{figure}[tbh]
\centering
\includegraphics[width=0.45\textwidth]{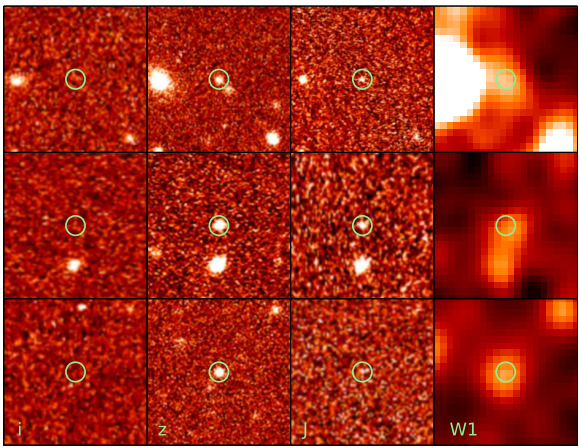}
\caption{The thumbnails of three confirmed quasars in each of $i$, $z$, J and W1 bands, in a region of $30\farcs0 \times 30\farcs0$ . The circles are in a radius of 2\farcs0. The postage stamps from upper to lower panels are J1048-0109, J1217+0131 and J1559+2212, respectively. Note that the resolution of WISE W1 is lower than that of optical and near-IR bands.}
\end{figure}

We also observed 23 $z\gtrsim6.5$ candidates and 19 $z \gtrsim 7$ candidates with Magellan/FIRE, and identified one quasar; DELS J104819.09$-$010940.21 (hereafter J1048$-$0109) at $z=6.63$. We used the same method mentioned above to measure the redshift. As our spectrum of J1048$-$0109 is of relatively low S/N, we estimated $m_{1450}$ by scaling a quasar composite spectrum \citep{vanden01} to the $J$-band magnitude of J1048$-$0109 and then fit a power-law continuum $f_\nu \sim \nu^{-0.5}$ to the composite spectrum. We thus derived an absolute magnitude of $M_{1450}=-25.95$. J1048$-$0109 is the first $z>6.5$ quasar discovered using DECaLS imaging. It was also independently discovered by Venemans et al. (2017, in preparation.) using VIKING imaging. Our low S/N and low resolution spectrum suggests that J1048$-$0109
is a quasar at $z\sim6.6$. The redshift derived using Mg\,{\sc ii} from a better spectrum is $z=6.661$ (B. Venemans, private communication) which is consistent with our measurement.
All of the other $z > 7$ candidates that we observed could either be classified as L/T dwarfs or had spectra without any flux break in the blue. For objects with no obvious blue-flux-break, we can only rule out quasars as an explanation---the spectra are of insufficient S/N to identify whether they are of stellar dwarfs or galaxies. The photometric information of all objects targeted for spectroscopy are listed in the APPENDIX. The thumbnails of three confirmed quasars in each of the selection bands are shown in Figure 5.

\section{Discussion and Summary}
The two new $z\sim6$ quasars we present are fainter than those in the SDSS main $z\sim6$  quasar sample \citep[e.g.][]{fan06b}, most of the Pan-STARRS1 $z\sim6$ quasar sample \citep[e.g.][]{banados16} and the break magnitude of the Quasar Luminosity Function (QLF) at $z\sim6$ \citep[e.g.][]{kashikawa15,jiang16}. Our discoveries demonstrate that DECaLS, together with near-IR imaging (e.g.\ from UHS) will be a highly promising dataset for finding more faint $z\sim6$ quasars. DECaLS thus has the potential to improve constraints on the QLF, particularly for the part fainter than the break magnitude. Using the $z\sim6$ QLF from \cite{jiang16}, we might expect to find $\sim1{,}000$ quasars at $z\sim6$ with $z$-band brighter than 23.0 (the projected final depth of DECaLS) and $\sim200$ quasars at $z\sim6$ with $z$-band brighter than 21.5 with DECaLS$+$UHS/ULAS over the entire DECaLS footprint.

DECaLS does not conduct deep $i$-band imaging, which is a limitation for finding $i$-dropouts. With forced photometry on SDSS images we can only reach $z\sim$21.5 since the 2-$\sigma$ magnitude of SDSS $i$-band is $\sim23.5$ \citep{gunn98}. In addition, current near-IR data from UKIDSS and UHS is insufficiently deep to target $z>21.5$ candidates. In future work, in order to further reject false $i$-dropouts and to push to fainter magnitudes due to the shallow SDSS $i$-band photometry, we plan to stack SDSS and Pan-STARRS1 $i$-band images for our quasar candidates. The stacked $i$-band images will allow us to push to $z\sim22.0$. Targeting fainter $z\sim6$ quasar candidates will require additional deep optical imaging to further eliminate cool dwarf contaminants. We will use the upgraded Wide-Field Camera at the Wyoming Infrared Observatory (WIRO) 2.3-meter classical Cassegrain telescope \citep{findlay16} for i-band imaging followups for our quasar candidates in order to further reject contaminations.

Although we have discovered one $z>6.5$ quasar with DECaLS data, the efficiency ($\sim$4\%) of our selection is quite low. In order to improve the efficiency, we will have to include additional imaging data, such as Pan-STARRS1 $y$-band, in order to further reject contaminants. The Pan-STARRS1 $y$-band is narrower than the typical near-IR $Y$ band, and will thererore be more effective in rejecting L/T dwarfs using $y-J$ colors. Several modern techniques, such as probabilistic selections based on Bayesian model or template fitting, have been used to search quasars at different redshift range \citep[e.g. ][]{mortlock12, matsuoka16}. We will combine these techniques with our color box selection to further improve the efficiency of selecting high redshift quasars in the future work. Combining UHS, Pan-STARRS1, DECaLS and upcoming BASS and MzLS imaging, we expect to be able to detect quasars up to $z>6.5$ over a large area ($\sim$12{,}000\,deg$^{2}$) at latitudes far from the Galactic plane. Based on extrapolating the $z\sim6$ QLF from \cite{jiang16} to higher redshift \citep{fan01b}, we expect to find $\sim30$ $z>6.5$ quasars over $\sim$12{,}000\,deg$^{2}$ to $J_{\rm VEGA}<19.5$.

\acknowledgments
We thank the anonymous referee for carefully reading the manuscript and providing constructive comments and suggestions.
F.W., J.Y. and X.-B.W. acknowledge support from NSFC grants 11373008 and 11533001, and the National Key Basic Research Program of China 2014CB845700. X.F. and I.D.M. acknowledge support from US NSF grant AST 15-15115. L.J. acknowledges support from the Ministry of Science and Technology of China under grant 2016YFA0400703. N.P.R acknowledges support from the STFC and the Ernest Rutherford Fellowship scheme. This research uses data obtained through the Telescope Access Program (TAP), which has been funded by the Strategic Priority Research Program ''The Emergence of Cosmological Structures'' (Grant No. XDB09000000), National Astronomical Observatories, Chinese Academy of Sciences, and the Special Fund for Astronomy from the Ministry of Finance. We acknowledge the use of the MMT Observatory, a joint facility of the Smithsonian Institution and the University of Arizona. We acknowledge the use of the Magellan telescope and ANU 2.3m telescope. We also acknowledge the use of public data, including ALLWISE, UKIDSS, VHS and VIKING.

{\it Facilities:} \facility{Magellan (FIRE)}, \facility{MMT (Red Channel spectrograph)}, \facility{2.3m/ANU (WiFeS)}.

\appendix
\begin{deluxetable*}{ccrrrrrrrrrrrrrrcrl}
\tabletypesize{\scriptsize}
\tablecaption{Photometric information for objects selected for spectroscopy.}
\tablewidth{0pt}
\tablehead{
\colhead{Name} & \colhead{$z_{\rm AB}$} & \colhead{$z_{\rm err}$} & \colhead{$J$}& \colhead{$J_{\rm err}$} & \colhead{$W1$} & \colhead{$W1_{\rm err}$} & \colhead{Flag\tablenotemark{a} } & \colhead{Instrument} & \colhead{Quasar}
}
\startdata
J155909.09+221214.43&20.60 &0.05 &19.86 &0.25 &17.14 &0.11 &z6&MMT/Red&YES\\
J121721.35+013142.52&20.89 &0.04 &20.21 &0.16 &17.58 &0.21 &z6&ANU/WiFes&YES\\
J080912.97+270609.39&21.37 &0.17 &19.33 &0.14 &17.54 &0.19 &z65&Magellan/FIRE&NO\\
J083236.21+052231.38&21.92 &0.24 &19.74 &0.16 &17.27 &0.16 &z65&Magellan/FIRE&NO\\
J085308.35+265015.81&21.67 &0.15 &19.56 &0.10 &17.67 &0.21 &z65&Magellan/FIRE&NO\\
J093701.69+105554.84&21.99 &0.17 &19.86 &0.13 &17.32 &0.17 &z65&Magellan/FIRE&NO\\
J103456.70+003329.87&21.90 &0.07 &19.72 &0.15 &17.65 &0.19 &z65&Magellan/FIRE&NO\\
J103900.49+062624.38&21.66 &0.11 &19.59 &0.13 &17.55 &0.19 &z65&Magellan/FIRE&NO\\
J104819.09-010940.21&21.95 &0.08 &19.71 &0.17 &17.34 &0.17 &z65&Magellan/FIRE&YES\\
J110943.88+053123.58&21.80 &0.21 &19.68 &0.15 &17.71 &0.22 &z65&Magellan/FIRE&NO\\
J111421.40+085419.84&21.69 &0.12 &19.85 &0.17 &17.26 &0.15 &z65&Magellan/FIRE&NO\\
J112025.67+045133.65&21.86 &0.13 &19.78 &0.18 &17.26 &0.16 &z65&Magellan/FIRE&NO\\
J113127.92+083826.59&22.14 &0.15 &20.01 &0.15 &17.51 &0.20 &z65&Magellan/FIRE&NO\\
J113909.45+090359.29&22.39 &0.27 &20.22 &0.17 &17.58 &0.22 &z65&Magellan/FIRE&NO\\
J122209.82+041730.64&21.68 &0.09 &19.50 &0.18 &17.17 &0.14 &z65&Magellan/FIRE&NO\\
J123809.62+143428.66&21.43 &0.15 &19.22 &0.14 &16.79 &0.08 &z65&Magellan/FIRE&NO\\
J130911.21+140725.15&21.77 &0.22 &19.76 &0.15 &17.42 &0.17 &z65&Magellan/FIRE&NO\\
J131738.06+261354.20&22.17 &0.31 &19.95 &0.19 &17.60 &0.15 &z65&Magellan/FIRE&NO\\
J132836.42+060154.78&21.46 &0.09 &19.43 &0.20 &17.80 &0.20 &z65&Magellan/FIRE&NO\\
J132932.03+134243.23&21.89 &0.16 &19.89 &0.16 &17.37 &0.13 &z65&Magellan/FIRE&NO\\
J134426.87+085951.35&22.32 &0.30 &20.24 &0.19 &17.74 &0.18 &z65&Magellan/FIRE&NO\\
J134553.48+104258.08&21.56 &0.11 &19.68 &0.19 &17.81 &0.19 &z65&Magellan/FIRE&NO\\
J150052.55+004705.35&21.83 &0.11 &19.64 &0.18 &17.30 &0.15 &z65&Magellan/FIRE&NO\\
J150404.95+092456.93&21.88 &0.16 &19.74 &0.14 &17.39 &0.13 &z65&Magellan/FIRE&NO\\
J155028.18+074457.91&21.60 &0.09 &19.65 &0.16 &17.63 &0.20 &z65&Magellan/FIRE&NO\\
J074244.36+294443.86&21.77 &0.15 &18.31 &0.06 &15.94 &0.06 &z7&Magellan/FIRE&NO\\
J075140.52+164159.00&22.14 &0.30 &19.08 &0.10 &17.26 &0.14 &z7&Magellan/FIRE&NO\\
J081019.36+102556.04&21.61 &0.20 &18.51 &0.10 &16.29 &0.06 &z7&Magellan/FIRE&NO\\
J092654.89+213852.38&21.66 &0.23 &18.61 &0.09 &15.95 &0.06 &z7&Magellan/FIRE&NO\\
J094645.96+261941.90&21.91 &0.36 &18.56 &0.12 &16.88 &0.11 &z7&Magellan/FIRE&NO\\
J101800.41+290434.94&21.59 &0.15 &18.58 &0.08 &16.33 &0.07 &z7&Magellan/FIRE&NO\\
J114246.78+275420.95&22.02 &0.30 &18.78 &0.10 &16.07 &0.06 &z7&Magellan/FIRE&NO\\
J132439.48+222151.84&22.18 &0.30 &18.93 &0.10 &16.93 &0.11 &z7&Magellan/FIRE&NO\\
J151545.16+282600.41&21.40 &0.11 &18.39 &0.07 &15.85 &0.11 &z7&Magellan/FIRE&NO\\
J155830.17+030052.35&21.67 &0.09 &18.24 &0.07 &16.30 &0.07 &z7&Magellan/FIRE&NO\\
J161013.32+132029.49&22.34 &0.61 &18.89 &0.09 &16.90 &0.10 &z7&Magellan/FIRE&NO\\
J161641.30+184509.36&22.39 &0.20 &19.25 &0.18 &17.02 &0.10 &z7&Magellan/FIRE&NO\\
J161947.08+001531.58&22.11 &0.16 &18.66 &0.09 &16.74 &0.10 &z7&Magellan/FIRE&NO\\
J162627.64+061222.10&21.38 &0.18 &18.36 &0.06 &15.68 &0.05 &z7&Magellan/FIRE&NO\\
J163508.35+224529.96&22.44 &0.26 &18.98 &0.13 &17.37 &0.12 &z7&Magellan/FIRE&NO\\
J164343.09+140320.27&21.57 &0.10 &18.39 &0.09 &16.14 &0.06 &z7&Magellan/FIRE&NO\\
J165746.46+003512.56&21.94 &0.13 &18.87 &0.09 &16.97 &0.11 &z7&Magellan/FIRE&NO\\
J170806.11+205553.58&22.66 &0.21 &19.33 &0.14 &17.10 &0.10 &z7&Magellan/FIRE&NO\\
J172208.00+243325.41&22.57 &0.25 &18.79 &0.13 &16.43 &0.06 &z7&Magellan/FIRE&NO

\enddata
\tablecomments{The exposure time of Magellan/FIRE longslit mode observations are around $5-10$ minutes. For objects with no obvious blue-flux-break, we can only rule out quasars as an explanation---the spectra are of insufficient S/N to identify whether they are of stellar dwarfs or galaxies.}
\tablenotetext{a}{This column flags the color boxes used to select candidates. Objects flagged as "z6", "z65" and "z7" are selected using the blue box, orange box and red box in Figure 1, respectively.} 

\end{deluxetable*}

\end{document}